\documentclass{caosp302}

\usepackage{graphicx}

\articleNo{123}
\pubyear{2007}
\volume{35}
\volnumber{3}
\firstpage{1}
\received{November 19, 2007}
\accepted{November 28, 2007}

\begin{document}

%
\hauthor{V. \,Khalack {\it et al.}}

\title{Vertical stratification of iron abundance in atmospheres of blue horizontal-branch stars}


%
\author{
        V.\,Khalack \inst{1,}
      \and
        F.\,LeBlanc \inst{1}
      \and
        B.B.\,Behr \inst{2}
      \and
        G.A.\,Wade \inst{3}
      \and
        D.\,Bohlender \inst{4}
       }

%
\institute{
           D\'epartement de Physique et d'Astronomie,
               Universit\'e de Moncton,\\
         \and
           Department of Astronomy, University of Texas at Austin\\
         \and
           Department of Physics, Royal Military College of Canada\\ 
         \and
           National Research Council of Canada,
               Herzberg Institute of Astrophysics
          }

\date{March 8, 2003}

\maketitle

\begin{abstract}
The observed slow rotation and abundance peculiarities of certain
blue horizontal branch (BHB) stars suggests that atomic diffusion
can be important in their stellar atmospheres and can lead to
vertical abundance stratification of chemical species in the atmosphere.
To verify this hypothesis, we have undertaken an
abundance stratification analysis in the atmospheres of six BHB
stars, based on recently acquired McDonald-CE spectra. Our numerical
simulations show that the iron abundance is vertically stratified in the
atmospheres of two stars in M15: B267 and B279. One star WF2-2541 in
M13 also appears to have vertically stratified iron abundance, while
for WF4-3085 the signatures of iron stratification are less
convincing. In all cases the iron abundances increase towards the
lower atmosphere. The other two stars in our sample, WF4-3485 and
B84, do not show any significant variation of iron with atmospheric
depth. Our results support the idea that atomic diffusion dominates
other hydrodynamic processes in the atmospheres of BHB stars.
\end{abstract}

%
\section{Introduction}
\label{intr}
Atomic diffusion is often proposed to explain the various observed anomalies
(abundance peculiarities, photometric jumps and gaps, and low gravities) of BHB stars.
Comprehensive surveys of BHB star abundances show
that stars hotter than $T_{eff}\simeq$ 11,500K have abundance anomalies as compared to
the other stars in the same globular cluster.
Behr et al. (2000) demonstrated that the BHB stars with $T_{eff}>$ 11,500K show modest
rotation ($v\sin{i}<$ 10~km/s), while the cooler stars are rotating more rapidly.


The metal abundance anomalies and slow rotation suggest that microscopic atomic diffusion
is effective in stellar atmospheres of BHB stars with $T_{eff}>$ 11,500~K. In this scenario, the
competition between radiative levitation (acting primarily through bound-bound atomic transitions) and
gravitational settling yields a net acceleration on atoms, which results in their diffusion
in the atmosphere. This process naturally produces vertical abundance stratification of
different chemical species. Direct estimation of this stratification from line profile analysis
would be a convincing argument in favour of efficient atomic diffusion in the atmospheres of
hot BHB stars.

\section{Results of spectral analysis}
The line profile simulations are performed in a Phoenix 
LTE stellar atmosphere model with solar metallicity, but with enhanced iron and
depleted helium abundances (Behr 2003), employing the {\sc Zeeman} spectrum synthesis code.
For each line profile the iron abundance, radial velocity and $v\sin{i}$ were fitted using
an automatic minimization routine (Khalack et al. 2007a).

All the analysed stars are slowly rotating objects and have strong He depletion.
The results of our numerical simulations show that iron appears to be vertically
stratified in the atmospheres of three stars: B267 and B279 in M15 and WF2-2541 in M13.
The Fe abundance  increases towards the lower atmosphere,
while for the upper atmospheric layers it is near its solar value. For WF4-3085 in M13 we
can not reach a final conclusion, because models with higher (by 1000 K) $T_{eff}$ and with
solar abundances do not provide confident results for iron stratification, taking into account
the uncertainties. The other two stars, WF4-3485 in M13 and B84 in M15, show no evidence of
stratification of iron. 
More details concerning these results are discussed by Khalack et al. (2007b.)

\section{Discussion}

We have found, for the first time, evidence of vertical stratification of iron in the
atmospheres of three BHB stars. These results support the common belief that atomic
diffusion is important in the atmospheres of these objects. As abundance stratification will
modify the atmospheric structure, such observational results can serve as constraints in the
development of atmospheric models such as those of Hui-Bon-Hoa et al. (2000).

\end{document}